\documentclass[10pt,prb,aps,twocolumn]{revtex4}

\usepackage[dvips]{graphicx}

\begin{document}
\title{Noise-induced effects in magnetization reversal and chirality control
of circular array of single-domained nanomagnets}

\author{A.L. Pankratov}
\email[]{alp@ipm.sci-nnov.ru}
\author{S.N. Vdovichev}
\author{I.M. Nefedov}
\author{I.R. Karetnikova}
\affiliation{Institute for Physics of
Microstructures of RAS, Nizhny Novgorod, 603950, Russia}

\begin{abstract}
The effect of noise on the process of high-speed remagnetization of vortex state of a pentagonal array of five circular magnetic nanoparticles is studied by means of computer simulation of Landau-Lifshits model. The mean switching time and its standard deviation of the reversal between the counterclockwise and clockwise vorticities have been computed. It has been demonstrated that with the reversal by the pulse with sinusoidal shape, the optimal pulse duration exists, which minimizes both the mean switching time (MST) and the standard deviation (SD). Besides, both MST and SD significantly depend on the angle between the reversal magnetic field and pentagon edge, and the optimal angle roughly equals 10 degrees. Also, it is demonstrated that the optimization of the angle, duration and the amplitude of the driving field leads to significant decrease of both MST and SD. In particular, for the considered parameters, the MST can be decreased from 60 ns to 2-3 ns. Such a chain of magnetic nanoparticles can effectively be used as an element of magnetoresistive memory, and at the temperature 300K the stable operation of the element is observed up to rather small size of nanoparticles with the raduis of 20 nm.
\end{abstract}

\maketitle

Studies on arrays of magnetic nanoparticles have arisen growing interest due to their potential applications in spintronics devices \cite{Hir,SD}, and especially in the past decade with the development of modern technologies, allowing to produce regular arrays of magnetic nanoparticles with minimal sizes of order 20 nm. The peculiarities of the magnetic collective behavior have essentially been investigated by numerical calculations taking into account the relevant interparticle interactions. One of the directions of such investigations is the study of the magnetic flux closure (MFC) state or the magnetization vortex curling state realized within a single magnetic nanoelement. Such artificially designed arrays of magnetic nanoparticles can have clear application for magnetic logic devices \cite{SD}-\cite{Pra}. On the other hand, one of perspective realizations of the magnetic random access memory (MRAM) uses magnetic rings with vortex-like distribution of magnetization. The interest to such vortex states is due to two reasons: first, the crosstalk fields are small in this state, allowing high density of elements. The second, the remagnetization of such elements can simultaneously be driven by magnetic field of a current, having vortex-like structure, and also with the use of the spin torque effect. However, considering large circular nanoparticles\cite{Gai}, it should be noted that it is practically impossible to perform their optimization in frequency response and stability to thermal noise due to large calculation time. That is why it is important to study fluctuational effects in systems of only few magnetic nanoparticles where MFC states are realized. In Ref. \cite{Kong} the cluster of five circular magnetic nanoparticles forming a ring (let us call it the pentagonal array) has been considered and the approach for the formation and manipulation of the chirality of a magnetization flux closure (MFC) state has been proposed. However, the study has been focused on the hysteretic loops and such interesting topic as switching times between opposite vortex states has not been investigated. Besides, the effect of thermal fluctuations on the stability of MFC state and switching times were not discussed.

Due to small size the nanoparticles can often be considered as single-domain particles \cite{thi}-\cite{sch4}, since the inhomogeneous distribution of the magnetization is energetically disadvantageous. Due to the complexity of the model, described by the time-dependent Landau-Lifshits equation with noise \cite{AA}, starting from the seminal paper \cite{brown} mostly the relaxation times of magnetization in the case of static magnetic field have been studied \cite{kalm95}-\cite{kalm98}.  Ref.s \cite{DenEPL,DenPRL} were focused on the effect of noise and high-frequency rotating magnetic field on the magnetic dipole dynamics, while Ref. \cite{Raik} was devoted to investigation of stochastic resonance effect. However, the practically interesting effect of fluctuations on the high-speed switching of magnetic single-domain particles has been investigated only recently \cite{noise},\cite{SP}.

In the present paper the effect of thermal fluctuations on the process of high-speed remagnetization of vortex state of a chain of five circular uniformly magnetized nanomagnets, located in vertices of a pentagon, is studied by means of computer simulation of Landau-Lifshits model. It is focused on the investigation of statistical characteristics of the reversal process with the aim to find an optimal regime of reversal between different vortex states with the smallest mean reversal time and the standard deviation.

\begin{figure}[h]
\resizebox{1\columnwidth}{!}{
\includegraphics{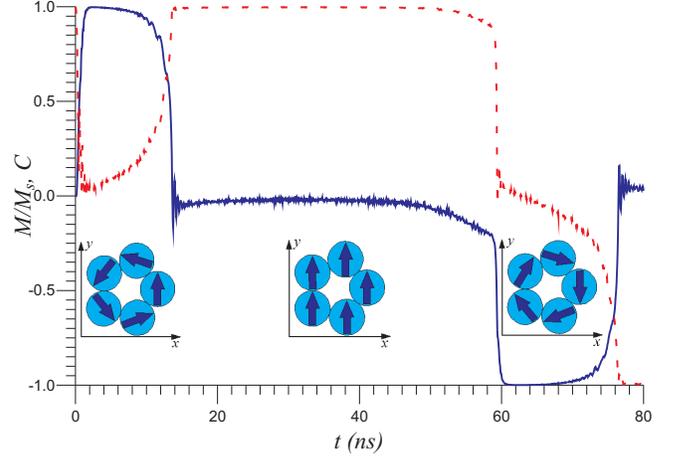}}
\caption{The temporal evolutions of magnetization (dashed curve) and chirality (solid curve). } \label{fig1}
\end{figure}

The dynamics of a magnetic nanoparticle of an array is described by the Landau-Lifshits equation \cite{AA}:
\begin{equation}
\frac{d\overrightarrow{M}}{dt}=-\frac{\gamma}{\beta}\left[\overrightarrow{M}\times
\overrightarrow{H}\right]-\frac{\alpha\gamma}{\beta M_s}
\left[\overrightarrow{M}\times\left[\overrightarrow{M}\times\
\overrightarrow{H}\right]\right], \label{LL}
\end{equation}
where $\overrightarrow{M}$ is the magnetization of a particle,
$\overrightarrow{H}$ is the effective magnetic field, $\gamma$ is
the gyromagnetic constant, $\alpha$ is the damping, $\beta=1+\alpha^2$, $M_s=\left|\overrightarrow{M}\right|$ is the saturation magnetization. The effective magnetic field contains the following components:
$\overrightarrow{H}=\overrightarrow{H_d}+\overrightarrow{H_e}+\overrightarrow{H_T}$,
where $\overrightarrow{H_d}$ is the field of magnetic-dipole interaction,
$\overrightarrow{H_e}$ is the external field, and
$\overrightarrow{H_T}$ - fluctuational field. The fluctuational
field is assumed to be white Gaussian noise with zero mean and the
correlation function:
$\left<H(t)_{Ti}H(t')_{Tj}\right>=\frac{2\alpha kT}{\gamma M_s
V}\delta(t-t')\delta_{ij}$, where $k$ -- Boltzmann constant, $T$ is
the temperature, and $V=\frac{4}{3}\pi(\lambda R)^3$ is the volume of the magnetic particle. Here, as in \cite{Kong},
each magnetic nanoparticle is assumed to have a core-shell structure, with a soft ferromagnetic core encapsulated
by a thin shell of nonmagnetic material. The core-shell structure is specified by two geometric parameters,
which are the outer radius $R$ and the core-to-particle radius ratio, $\lambda=R_{core}/R$, where $R_{core}$ is the radius of the magnetic
core. Adjacent nanoparticles in the array are physically in contact to form a compact array structure so that the outer
diameter of each particle $2R$ is the particle spacing. Since the magnetic core is made of soft ferromagnetic material in perfect
spherical shape, both the shape and the magnetocrystalline anisotropies are small and are not accounted for in the calculation. Owing to the presence of the nonmagnetic shell, the interparticle interaction is limited only to the long-range dipolar magnetic interaction. Below we will study two cases $\lambda=0.9$ and $\lambda=1.0$.
The magnetic-dipole interaction field for the $i$-th dipole has the following form:
\begin{equation}
\overrightarrow{H}_{di}=V_R\sum\limits_{j\ne i} \left[\frac{3\left(\overrightarrow{M}_j\overrightarrow{r}_{ij}\right)\overrightarrow{r_{ij}}}{r_{ij}^5}-\frac{\overrightarrow{M}_j}{r_{ij}^3}\right], \label{Hd}
\end{equation}
where $V_R=\frac{4}{3}\pi(\lambda R)^3/R^3$, $\overrightarrow{r}_{i}$ is the radius-vector of $i$-th point, $r_{ij}=\left|r_i-r_j\right|$ is the distance between positions $i$ and $j$ in units of the particle radius $R$, $(\overrightarrow{r}_{ij})_x=\frac{1}{\sin(\pi/N)}\left(\cos\frac{2\pi i}{N}-\cos\frac{2\pi j}{N}\right)$, $(\overrightarrow{r}_{ij})_y=\frac{1}{\sin(\pi/N)}\left(\sin\frac{2\pi i}{N}-\sin\frac{2\pi j}{N}\right)$ for $N$=5 in our case.
If the switching magnetic field is absent, due to magnetic-dipole interaction, the vorticity will be formed almost instantly, see Fig. \ref{fig1}. Its direction depends on the symmetry of the system of dipoles and it is possible to control the sign of the vorticity using uniform external magnetic field, exceeding the interaction field. For example, if the initial state is counterclockwise (see the left inset of Fig. 1), and an external field is applied along $y$-axis and along the pentagon edge, the magnetic state becomes parallel to the $y$-axis (see the central inset of Fig. 1). After switching off the magnetic field, the vorticity will be clockwise, see the right inset of Fig. \ref{fig1}.
To describe the vorticity quantitatively, one can introduce the chirality similarly to Ref. \cite{Gai}:
\begin{equation}
C=\frac{1}{N M_s R}\sum\limits_{i} \left(x_i M_{yi} -y_i M_{xi}\right), \label{chi}
\end{equation}
where $x_i=\cos(2\pi i/N)$ and $y_i=\sin(2\pi i/N)$ if the coordinate origin is chosen at the center of the pentagon, $M_{xi}$ and $M_{yi}$ are $x$ and $y$ projections of magnetization of the $i-th$ particle, respectively.

To find the area of parameters where the fastest and the most reliable reversal occurs, as the characteristic to be studied let us choose the first passage time of a certain boundary, e.g., when the chirality, starting from the positive value, see Fig. \ref{fig1}, crosses its -1/2 value level. The mean first passage time (the mean reversal time, MRT) $\tau$, and the standard deviation of the first passage time $\sigma$ (SD, jitter) are \cite{ACP}: $\tau$=$\langle t \rangle$=$\sum_{i=1}^{K}t_i/K$, $\langle t^2 \rangle$=$\sum_{i=1}^{K}t_i^2/K$, $\sigma$=$\sqrt{\langle t^2\rangle-\langle t \rangle^2}$, where $t_i$ is the first passage time of an absorbing boundary and $K\ge 10000$ is the number of realizations.

In the calculations \cite{calc} let us choose the same parameters as in \cite{Kong}: $\alpha$=$0.1$,
$\gamma$=$1.76\cdot 10^{7}{\rm Hz/Oe}$, $M_s$=1400 ${\rm Oe}$. For modeling we take different amplitudes of the magnetic field
in the region from $H_0$=1000 ${\rm Oe}$ to $H_0$=2800 ${\rm Oe}$, such that the field $H_0$=1000 ${\rm Oe}$ slightly exceeds the magnetic dipole interaction field. In the following we will study different radii of the particles, starting from 10nm to 30nm.
As an example of a driving with smooth fronts we consider the sinusoidal pulse $\vec{H_e}$=$\vec{e}H_0\sin\pi t/t_p$ with the width $t_p$, where $\vec{e}$ is the unitary vector of magnetic field direction. If the switching during $t_p$ does not happen, the computation is continued for $\overrightarrow{H_e}$=0 until some long period of time $t_f$, much larger than any other relaxation time scale. Our aim is to find the parameter range, for which the reversal by a smooth pulse takes place quickly and reliably.

\begin{figure}[h]
\resizebox{1\columnwidth}{!}{
\includegraphics{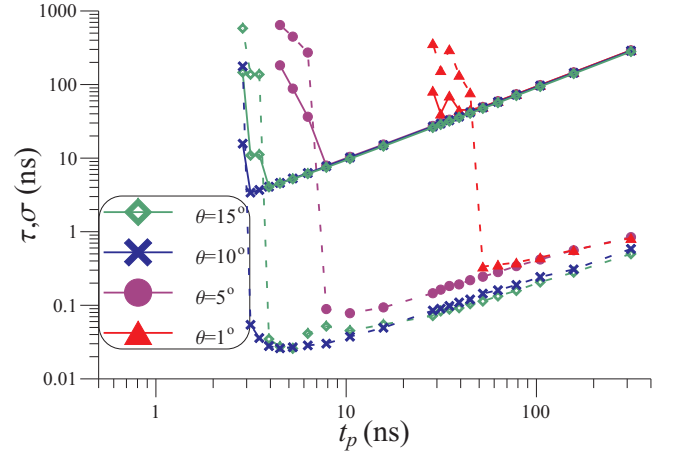}}
{\caption{The MRT (solid curves) and SD (dashed curves) versus pulse
width for different angles between the pentagon edge and external
magnetic field, the temperature $T$=300 K. }
\label{fig2}}
\end{figure}
Let us start our study examining the dependence of the MRT and SD versus the angle $\theta$ between the pentagon edge and the driving field.
Presence of fluctuations or small asymmetry of magnetic field will lead to a fast formation of the vorticity (\ref{chi}), see Fig. \ref{fig1} for $\theta=1^\circ$ and $T=300$ K, where the initial magnetization was uniform as in central inset of Fig. \ref{fig1}, and the vorticity was formed in 1 ns. In Fig. \ref{fig2} the MRT and SD are presented versus the pulse duration for different angles between the pentagon edge and external magnetic field. One can see that this dependence is very similar to the single dipole case, compare with Ref. \cite{noise}: both MRT and SD demonstrate minima as functions of the pulse width and there is a strong dependence on the angle $\theta$. The decrease of temporal characteristics at large $t_p$ is due to the fact that with decrease of the pulse width the potential barrier disappears faster. With further shortening of the pulse, the magnetization does not have enough time for the complete reversal during $t_p$, so the MRT increases. In particular, the reversal time can not be smaller than the formation of vorticity, which is of order of 1 ns, see Fig. \ref{fig1}. This, actually, means that for rather short pulses the transition occurs due to effect of fluctuations (the so-called noise-induced switching). The same applies to SD, since for smaller switching time the noise has less possibility to move the system far from the deterministic trajectory, existing in the absence of noise. The existing of the optimal driving angle is again similar to a single dipole case and is explained by existence of an unstable state of the system, leaving which seriously delays the process of switching. One can see that changing the driving angle from $\theta=1^\circ$ (whose value is close to the unstable state $\theta=0^\circ$) to $\theta=15^\circ$ allows to speedup the switching process by more that one order of magnitude, and to correspondingly reduce the SD. However, in difference with the single dipole case, the optimal angle here is not $\theta=45^\circ$ \cite{tilt}, \cite{perp}, \cite{noise}, but roughly $\theta=10^\circ$, which is 1/8 of the angle between two subsequent dipoles in the pentagonal array.

\begin{figure}[h]
\resizebox{1\columnwidth}{!}{
\includegraphics{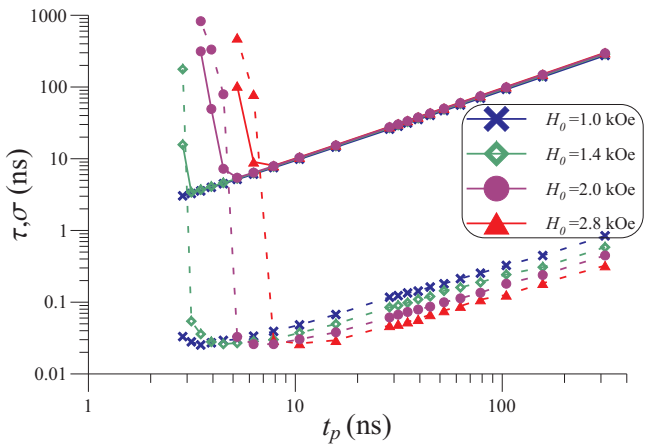}}
{\caption{The MRT (solid curves) and SD (dashed curves) versus the
pulse width for different driving pulse amplitude $H_0$, the temperature $T$=300 K, $\theta=10^\circ$. } \label{fig3}}
\end{figure}
An unusual situation in comparison with a single dipole case occurs if one tries to study the dependence versus the driving field amplitude. This dependence is presented in Fig. \ref{fig3} for the optimal driving angle $\theta=10^\circ$. Here, larger driving field delays the switching. For example, the increase of the field amplitude from $H_0$=1400 ${\rm Oe}$ to $H_0$=2800 ${\rm Oe}$ increases the minimal MRT in almost two times. Nevertheless, the minimal attainable SD is nearly the same and is proportional to $\sqrt{T}$ as it must.

\begin{figure}[h]
\resizebox{1\columnwidth}{!}{
\includegraphics{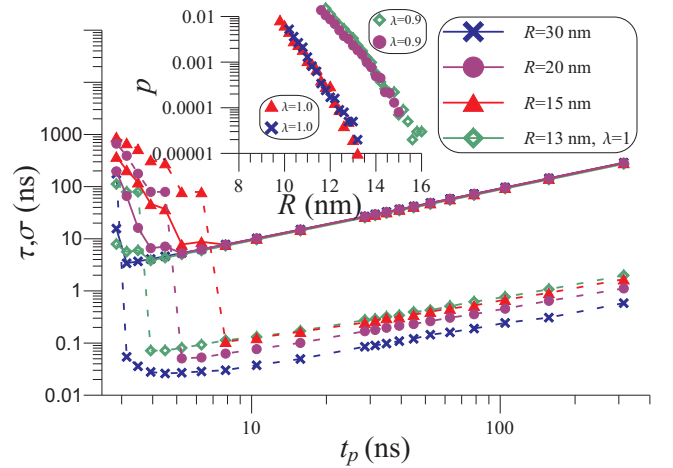}}
{\caption{The MRT (solid curves) and SD (dashed curves) versus the pulse width for different values of the particle radius and $\lambda$ for the temperature $T$=300 K, $\theta=10^\circ$. Inset: the nonswitching probability versus particle radius for different $\lambda$ and $t_p$} \label{fig4}}
\end{figure}

Finally, let us study the dependence from the radius of the particle, see Fig. \ref{fig4}, which is very important to understand the potential applications of such vortex states as components for MRAM. Note that, due to the normalizations chosen, the only dependence of the noise intensity on the particle radius remains. As one can see from Fig. \ref{fig4}, formally, even for $R=15$ nm for the given parameters the reversal still occurs stably. To check this more precisely, let us perform the simulation of nonswitching probability, which for many important applications must be below $10^{-15}$. Such low probabilities can not be computed numerically due to enormous calculation time. Let us perform simulations for $10^{5}$ realizations, and calculate the nonswitching probabilities $p(R)$ for different pulse durations as functions of the particle radius. The results of simulations are presented in the inset of Fig. \ref{fig4}. For $\lambda=0.9$ we choose $t_p$=10 ns and $t_p$=20 ns, while for $\lambda=1.0$ -- $t_p$=5 ns, and $t_p$=10 ns. As one can see from the inset of Fig. \ref{fig4}, the results $p(R)$ for different $\lambda$ seriously differ, while the results for different $t_p$, but the same $\lambda$, actually agree. It is clear from the presented digits and linear fit of the curves that even for $\lambda=1.0$ the probability $10^{-12}$ is reached for $R\approx20$ nm. While the considered parameters range corresponds to the regime with significant impact to noise, and below $10^{-5}$ the nonswitching probability can in fact decrease faster, it is recommended to use magnetic particles with larger $M_s$ values, and to increase the magnetic radius up to the particle radius, $\lambda \to 1$, in order to decrease the particle radius to 20 nm.

\begin{figure}[h]
\resizebox{1\columnwidth}{!}{
\includegraphics{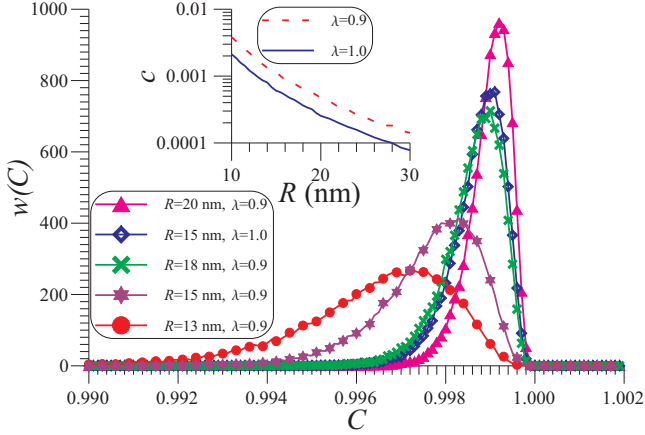}}
{\caption{The distribution function of chirality $C$ for different values of the particle radius and $\lambda$ for the temperature $T$=300 K. Inset: plot of $c=\sqrt{\left<C^2\right>-\left<C\right>^2}/\left<C\right>$.} \label{fig5}}
\end{figure}

To understand the reason of rather large non-switching probability, let us investigate the distribution function $w(C)$ of chirality for different values of the particle radius and $\lambda$ for the temperature $T$=300 K. One can see from Fig. \ref{fig5} that the distributions are rather broad and the distribution for $R=15$ and $\lambda=1$ is more narrow than even for $R=18$ and $\lambda=0.9$, while equal magnetic radius would be for $R=16.7$ and $\lambda=0.9$, which is presumably due to stronger dipole interaction in this case. Introducing relative width of the distribution function $c=\sqrt{\left<C^2\right>-\left<C\right>^2}/\left<C\right>$, one can see from the inset of Fig. \ref{fig5} that the increase of $\lambda$ from 0.9 to 1.0 decreases the width of the distribution $w(C)$ in nearly two times. Therefore, the rather large non-switching probability is due to noise-induced instability of the vorticity even in the absence of the switching signal.

In the present paper the effect of noise on the process of high-speed remagnetization of vortex state of a pentagonal chain of five circular magnetic nanoparticles is studied by means of computer simulation of Landau-Lifshits model. The mean switching time and the standard deviation of the reversal between the vorticities have been computed. It has been demonstrated that with the reversal by the pulse with sinusoidal shape, the optimal pulse duration exists, which minimizes both the mean switching time (MST) and the standard deviation (SD). Besides, both MST and SD significantly depend on the angle between the reversal magnetic field and the pentagon edge, and the optimal angle equals $\theta\approx 10^\circ$. Also, it is demonstrated that the optimization of the angle, duration and the amplitude of the driving field leads to significant decrease of both MST and SD. In particular, for the considered parameters, the MST can be decreased from 60 ns to 2-3ns. Such a chain of magnetic nanoparticles can effectively be used as an element of magnetoresistive memory, and at the temperature 300K the stable operation of the element is observed up to rather small size of nanoparticles with the raduis of 20 nm for the chosen experimental parameters.

The authors wish to thank I.A. Shereshevskii and A.A. Fraerman for discussions. The work has been supported by RFBR-Povolzhie (project 08-02-97033).


\begin{thebibliography}{99}
\bibitem{calc}
The Landau-Lifshits equation with noise has been computed both by the Heun method programmed in Fortran and by the specialized package SIMMAG (Simulation of MicroMAGnets), developed in the Laboratory of Mathematical Modeling of IPM RAS.
\end{thebibliography}
\end{document}